\documentclass[a4paper,11pt]{amsart}

\usepackage{url}
\usepackage{cite}

\usepackage{amssymb}
\usepackage{color}

\usepackage[normalem]{ulem}
\usepackage[utf8]{inputenc}
\usepackage{graphicx}

\usepackage{soul}

\newcommand{\ourN}{{}^{n-1}\mathcal{N}}

\newcommand{\ourell}{{}\ell}

\newcommand{\qedskip}{\hfill $\Box$ \medskip}

\newcommand{\zP}{{\mathring{P}}}

\newcommand{\znabla}{{\mathring\nabla}}

\newcommand{\bean}{\begin{eqnarray}\nn}

\newcommand{\nn}{\nonumber}

 \newcommand{\mcV}{{\mycal V}}
 \newcommand{\GmcV}{ {G_\mcV}}
\newcommand{\GW}{ {G_W}}

\newcommand{\zV }{{ \mathring{V}}}
\newcommand{\zglorentz}{{ \mathring{\glorentz}}}
\newcommand{\glorentz}{{ {\mathbf g}}}

\newcommand{\zDelta}{{ {\mathring \Delta}}}
\newcommand{\zRriem}{{ {\mathring \Rriem}}}

\newcommand{\Rriem}{{ {\mathfrak R}}}
\newcommand{\RHriem}{{ {\mathfrak R}^H}}
\newcommand{\griem}{{ {\mathfrak g}}}
\newcommand{\Mriem}{{ {\mathfrak M}}}
\newcommand{\zgriem}{{ \mathring{\mathfrak g}}}

\DeclareFontFamily{OT1}{rsfs}{}
\DeclareFontShape{OT1}{rsfs}{m}{n}{ <-7> rsfs5 <7-10> rsfs7 <10->
rsfs10}{} \DeclareMathAlphabet{\mathscr}{OT1}{rsfs}{m}{n}
\newcommand{\mcM}{\mathscr M}

\newcommand{\eq}[1]{\eqref{#1}}

\newcommand{\bel}[1]{\begin{equation}\label{#1}}
\newcommand{\beal}[1]{\begin{eqnarray}\label{#1}}
\newcommand{\beadl}[1]{\begin{deqarr}\label{#1}}
\newcommand{\eeadl}[1]{\arrlabel{#1}\end{deqarr}}
\newcommand{\eeal}[1]{\label{#1}\end{eqnarray}}
\newcommand{\eead}[1]{\end{deqarr}}
\newcommand{\eea}{\end{eqnarray}}
\newcommand{\eeaa}{\end{eqnarray*}}

\newcommand{\be}{\begin{equation}}
\newcommand{\ee}{\end{equation}}

\DeclareFontFamily{OT1}{rsfs}{}
\DeclareFontShape{OT1}{rsfs}{m}{n}{ <-7> rsfs5 <7-10> rsfs7 <10->
rsfs10}{} \DeclareMathAlphabet{\mycal}{OT1}{rsfs}{m}{n}

\newcommand{\mcL}{{\mycal L}}
\newcounter{mnotecount}[section]

\newcommand{\N}{{\mathbb N}}

\newcommand{\rmnote}[1]{}%{\mnote{#1}}

\def\mysavedown#1{\edef\mysubs{\mysubs#1}}
\def\mysaveup#1{\edef\mysups{\mysups#1}}
\def\mydown#1{{\mytensor}_{\vphantom{\mysubs}#1}}
\def\myup#1{{\mytensor}^{\vphantom{\mysups}#1}}
\def\tensor#1#2{
  #1
  \def\mytensor{\vphantom{#1}}
  \def\mysubs{\relax}
  \def\mysups{\relax}
  \let\down=\mysavedown
  \let\up=\mysaveup
  #2
  \let\down=\mydown
  \let\up=\myup
  #2
  }

\newcommand{\Tr}{\operatorname{Tr}}

\newcommand{\C}{\mathbb C}
\newcommand{\R}{\mathbb R}

\renewcommand{\setminus}{\smallsetminus}

\renewcommand{\to}{\rightarrow}

\renewcommand{\epsilon}{\varepsilon}
\renewcommand{\hat}{\widehat}

\def\crn#1#2{{\vcenter{\vbox{
        \hbox{\kern#2pt \vrule width.#2pt height#1pt
           }
          \hrule height.#2pt}}}}

\renewcommand{\hbar}{{\overline h}}

\newcommand{\pre}[2]{{{\vphantom{#2}}^{#1}}\kern-.2ex{#2}}

\theoremstyle{plain}
\newtheorem{theorem}{\sc Theorem}[section]
\newtheorem{lemma}[theorem] {\sc Lemma}
\newtheorem{proposition}[theorem]{\sc Proposition}

\theoremstyle{definition}
\newtheorem{remark}[theorem]{\sc  Remark\rm}

\numberwithin{equation}{section}

\date{\today}

\begin{document}

\title[Stationary  spacetimes with negative $\Lambda$] {Non-singular spacetimes with a negative cosmological constant: IV.  Stationary  black hole solutions with matter fields}
\thanks{Preprint UWThPh-2017-19}

\author[P.T. Chru\'sciel]{Piotr T.~Chru\'sciel}

\address{Piotr
T.~Chru\'sciel, Faculty of Physics and Erwin Schr\"odinger Institute, University of Vienna, Boltzmanngasse 5, A1090 Wien, Austria}
\email{piotr.chrusciel@univie.ac.at} \urladdr{http://homepage.univie.ac.at/piotr.chrusciel/}

\author[Delay]{Erwann
Delay} \address{Erwann Delay, Universit\'e d'Avignon, Laboratoire de Math\'ematiques d'Avignon (EA 2151), 301 rue Baruch de Spinoza,
F-84916 Avignon, France}
\email{Erwann.Delay@univ-avignon.fr}
\urladdr{http://www.math.univ-avignon.fr/}

\author[P. Klinger]{Paul Klinger}

\address{Paul Klinger, Faculty of Physics and Erwin Schr\"odinger Institute, University of Vienna, Boltzmanngasse 5, A1090 Wien, Austria}
\email{paul.klinger@univie.ac.at}

\begin{abstract}
We use an elliptic system of equations with complex coefficients for a set of complex-valued tensor fields as a tool to construct infinite-dimensional families of non-singular stationary
black holes, real-valued Lorentzian solutions of the
Einstein-Maxwell-dilaton-scalar fields-Yang-Mills-Higgs-Chern-Simons-$f(R)$
equations with a negative cosmological constant. The families include  an infinite-dimensional family of solutions with the usual AdS conformal structure at conformal infinity.
\end{abstract}

\maketitle

\tableofcontents
\section{Introduction}
 \label{section:intro}

%\ptc{reread globally on 12 VIII 17}
There is currently considerable interest in the literature in space-times with a negative cosmological constant. This is fueled on one hand by studies of the AdS-CFT conjecture and of the implications thereof:
Indeed, this problem is of immediate physical interest in the context of the  weakly coupled supergravity limit of the  AdS/CFT correspondence.
On the other hand, these solutions are interesting because of a rich dynamical morphology: existence of periodic or quasi-periodic solutions, and of instabilities. All this leads naturally to the question of existence of stationary solutions of the Einstein equations with $
\Lambda<0$, with or without sources, and of properties thereof.

This manuscript is the fourth in a series of papers, starting with \cite{ChDelayStationary}, which are devoted to proving  existence of a large class of solutions to the Einstein equations with negative cosmological constant by perturbation of
known ones.   All of these papers are further related by the  fact that the field equations can be transposed to an elliptic system on a conformally compact {\it Riemannian} manifold. The  system is solved by an implicit function theorem argument under a non degeneracy hypothesis. This can be traced back to earlier work
of Graham and Lee~\cite{GL} on constructing Einstein metrics on $(n + 1)$-dimensional
balls with $S^n$ boundary, as generalized to more general infinities by Lee~\cite{Lee:fredholm} (compare~\cite{BiquardContinuation} for more general symmetric spaces).
Using such methods, in~\cite{ChDelayEM,ChDelayKlinger} we have constructed infinite dimensional families of  non-singular \emph{strictly stationary}
space times, solutions of the Einstein  equations with a negative cosmological constant and with various matter sources. These families include  an infinite-dimensional family of solutions with the usual AdS conformal structure at conformal infinity.
The construction there did not provide any black hole solutions, as strict stationarity is incompatible with existence of horizons. However, black hole solutions are of special interest. In fact, various such solutions have already been constructed numerically: For example, static Einstein-Yang-Mills black holes have been constructed in space-time dimension five in~\cite{WinstanleyEYM}, with four-dimensional solutions constructed in~\cite{BjorakerHosotani}, and higher dimensional ones in~\cite{RaduTchrakian}.  In~\cite{OkuyamaMaeda} an explicit five-dimensional such solution has been given. Rotating Einstein-Maxwell-Chern-Simons solutions have been presented in~\cite{BBSLMR}.
In~\cite{DiasHorowitzSantos} a family of five-dimensional black holes was constructed satisfying the Einstein-complex scalar field equations, with a stationary geometry and time-periodic scalar field; compare~\cite{AstefaneseiRadu}.
% \ptcr{~\cite{BML} don't do black holes, only solitons}
% \ptc{what about~\cite{IizukaIshibashiMaeda}? but we ignore dimension three, maybe we should not?}

The object of this work is to provide a rigorous existence proof for large families of such solutions.
The idea is to use a ``Wick rotation'' to construct suitable solutions of a system of \emph{elliptic equations with complex coefficients for a complex valued ``Riemannian metric''}. In a nutshell, we show that Lee's theorem on existence of perturbed Poincar\'e-Einstein Riemannian metrics~\cite[Theorem~A]{Lee:fredholm} can be extended to complex valued ``metrics'', and to more general equations, and that this can be used to construct stationary Lorentzian black hole solutions with large classes of matter sources. This proceeds as follows:

We wish to construct a Lorentzian metric ${}\glorentz $ in any
space-dimension $n\geq 3$,
with Killing vector $X=\partial/\partial t$, satisfying the Einstein-Maxwell-Chern-Simons-Yang-Mills-dilaton-scalar fields equations, with a stationary geometry but possibly time-periodic complex fields,   or the $f(R)$ equations.
In adapted coordinates the metric  can
be written as
\beal{8III17.1}
 &\glorentz  = -V^2(dt+\underbrace{\theta_i
 dx^i}_{=:\theta})^2 + \underbrace{g_{ij}dx^i dx^j}_{=:g}\, ,
 &
\\
 &
 \partial_t V = \partial_t \theta = \partial_t g=0\, .
 &
\eeal{8III17.2}

Let us introduce a \emph{complex parameter} $a\in \C$ and consider the complex-valued tensor field
\beal{20XI16.22}
 &\glorentz  = -V^2(dt+ a \theta)^2 +g \, ,
 &
\eea
satisfying \eqref{8III17.2}.
We will  say that a complex valued symmetric tensor field $g$ is a \emph{complex metric} if $g$ is symmetric and invertible.  Replacing $dt$ by $-i\, dt$  in \eqref{20XI16.22}, where $i=\sqrt{-1}$, we obtain a complex metric with Riemannian real part:
\beal{20XI16.23}
 & \griem:=   V^2(dt+ a i \theta)^2 +g \, .
 &
\eea
Under such a substitution the field equations transform in a controlled way, for example $t$-independent vacuum metrics lead to $t$-independent, possibly complex valued, tensor fields satisfying the vacuum equations, etc.

Working near a (real-valued) static Einstein metric $\mathring \griem=\mathring V^2 dt^2 + \mathring g$ satisfying a non-degeneracy condition  (as defined in the paragraph after Equation~\eq{10III17.4} below)  we will

\begin{enumerate}
  \item
 construct complex metrics reminiscent of \eq{20XI16.23} which solve the vacuum Einstein equations for small $|a|$, and
 \item show that $V$, $g$ and $\theta$ are real-valued if $a\in \R$.
\end{enumerate}

(Incidentally, we will also show that $\griem$, and hence $\glorentz$, is analytic in $a$,  an interesting property of the stationary metrics at hand which has does not seem to have been noticed so far.)
After ``undoing the Wick rotation'' leading from \eq{20XI16.22} to \eq{20XI16.23}, we will  show that the resulting Lorentzian space-time has a smooth event horizon at suitable zeros of $V$.

The construction guarantees that $V$ has zeros when $\mathring V$ did,  and leads indeed to the desired Lorentzian black-hole solution of the Einstein, or Einstein-matter equations.

Our non-degeneracy condition is satisfied by  large classes of metrics, including all four-dimensional Kottler metrics except the spherical ones with a single critical value of the mass parameter\cite{ChDelayKlingerNonDegenerate} (see also~\cite[Proposition~D.2]{ACD2}).  It is clear that the method opens further possibilities, which remain to be explored. For example, the technique is used in~\cite{ChDelayKlingerBoson} to construct boson star solutions.

%
%\red{
% \ptc{torus? say about spherical?}
%In an Appendix we show that our non-degeneracy condition is satisfied by the four-dimensional Schwarzschild-anti de Sitter metrics.%
%\footnote{It is conceivable that this result can be inferred from~\cite{IshibashiKodamaStability}. However, this is not clear, as the global conditions there do not seem to translate in any straightforward manner to those needed for our argument.}
%}

\section{Elliptic equations with complex principal symbol}
 \label{s19XI16.1}

Consider an $n$-dimensional (real) manifold $M$. Complex-valued tensor fields over $M$ are defined as sections of the usual (real) tensor bundles over $M$ tensored with $\C$. In other words, all coordinate transformations are real but we allow  tensors to have complex components. We emphasise that the ``Wick rotation'' above is \emph{not} considered to be a coordinate transformation, but a useful device mapping one set of equations and fields to another, more convenient, one.

As already mentioned, we will say that a two-covariant complex valued tensor $g$ is a \emph{complex metric} if $g$ is symmetric and invertible.

Let ${\Phi}=({\Phi}^A) $, $A=1,\ldots N$, be a collection of complex valued fields, forming a section  of a complex bundle over $M$. Let $g$ be a complex metric and consider a collection of $N$ equations
of the form
\bel{14I17.1}
 g^{i j} \partial_i \partial_j {\Phi}^A = F^A(g, \partial g, {\Phi}, \partial {\Phi})
 \,,
\ee
with some functions $F^A$ which will be assumed to depend smoothly upon their arguments.
This can be rewritten as the following collection of $2N$ equations for $2N$ real fields $(\Re {\Phi},\Im {\Phi})$:
\beal{14I17.2}
 \Re g^{i j} \partial_i \partial_j \Re {\Phi}^A
  & = & \Re \big( F^A(g, \partial g, {\Phi}, \partial {\Phi})\big) + \Im g^{i j} \partial_i \partial_j \Im {\Phi}^A
 \,,
\\
 \Re g^{i j} \partial_i \partial_j \Im {\Phi}^A
  & = & \Im \big( F^A(g, \partial g, {\Phi}, \partial {\Phi}) \big) - \Im g^{i j} \partial_i \partial_j \Re {\Phi}^A
 \,.
\eeal{14I17.3}
We will say that \eq{14I17.1} is elliptic if the system
\eq{14I17.2}-\eq{14I17.3} is elliptic in the usual sense for  PDEs involving real-valued functions.
The principal symbol of \eq{14I17.2}-\eq{14I17.3} is bloc-diagonal, built out of blocs of the form
\bel{2IV17.1}
 \left(
 \begin{array}{cc}
   \Re g^{i j} k_i k_j &- \Im g^{i j} k_ik_j  \\
  \Im g^{i j} k_i k_j &  \Re g^{i j} k_ik_j
 \end{array}
 \right)
 \,.
\ee
This is an isomorphism for $k\ne 0$ if and only if
\bel{2IV17.2}
(\Re g^{i j} k_i k_j)^2 +  ( \Im g^{i j} k_ik_j )^2 >0
 \,.
\ee
Hence, if $\Re g^{ij}$ is positive-definite  then \eq{14I17.1} will be elliptic regardless of $\Im g^{ij}$. More importantly for us, when $\Im g^{ij}$ is small enough  all the usual elliptic estimates, as needed for our analysis below,  apply to \eq{14I17.2}-\eq{14I17.3}, and hence to \eq{14I17.1}. Likewise, isomorphism properties for a real-valued $g$ carry over to nearby complex-valued $g$'s. As we will be using an implicit function theorem around real valued Riemannian metrics, our perturbation of $\Re g_{ij}$,  as well as $\Im g^{ij}$ will always be sufficiently small for the estimates and the isomorphism properties to remain valid.

\section{The setup}
 \label{s8III17.0}

We work in space-time dimension $d:=n+1$ and we normalise the cosmological constant to
\bel{10III17.21}
 \Lambda=-\frac{n(n-1)}2
  \,;
\ee
this can always be achieved by a constant rescaling of the metric.

Let $\znabla$ denote the covariant derivative associated with the metric $\zgriem$, set
\bel{10III17.2}
 \lambda^\mu:= \frac 1 {\sqrt{\det \griem}} \znabla_\alpha ({\sqrt{\det \griem}}  \griem^{\alpha \mu})
 \,.
\ee
(In \eq{10III17.2} the derivative $\znabla $ is of course understood as a covariant derivative operator acting on tensor densities.)
Denoting by $\Rriem^\mu{}_{\alpha\beta\gamma}$ the Riemann tensor of $\griem$, similarly for the Ricci tensor, we set
\bel{10III17.1}
 \RHriem_{\alpha\beta}:= \Rriem_{\alpha\beta}
  +
  \frac 12 \big(
 \griem _{\alpha\mu} \znabla_\beta \lambda^\mu
 +
 \griem_{\beta\mu} \znabla_\alpha \lambda^\mu
  \big)
 \,.
\ee
Then the linearisation with respect to the metric, at $\griem= \zgriem$, in dimension $d=n+1$, of the map
$$
 \zgriem \mapsto \Rriem^H_{\alpha\beta} +  (d-1) \griem_{\alpha\beta}
$$
is the operator
\bel{10III17.5}
 \zP:=\frac 12(\zDelta_L+2n)
 \,,
\ee
where the \emph{Lichnerowicz Laplacian} $\zDelta_L$  acts on   symmetric
two-tensor fields $h$ as
\bel{10III17.4}
 \zDelta_Lh_{\alpha \beta}:=-\znabla^\gamma\znabla_\gamma h_{\alpha \beta}
  +\zRriem_{\alpha \gamma }h^\gamma {_\beta}+\zRriem_{\beta\gamma }h^\gamma {_\alpha }-2\zRriem_{\alpha \gamma \beta \delta }h^{\gamma \delta}
  \,.
\ee
We will say that a metric $\zgriem$ is {\it non-degenerate} if $\zDelta_L+2n$ has no
$L^2$-kernel. This should not be confused with the notion of \emph{non-degenerate black holes}, also called \emph{extreme black holes}, which is the requirement of non-zero surface gravity.

Large classes of non-degenerate Einstein
metrics are described in~\cite{Lee:fredholm,ACD2,mand1,mand2}, see also Remark~\ref{R24III17.1} below.

It  follows immediately from the openness of the set of invertible operators  that if $\zgriem$ is a real-valued non-degenerate Riemannian metric, then all nearby (in a suitable topology, as determined by the problem at hand) complex valued metrics will also be non-degenerate.

The following is well-known (cf., e.g., the proof of theorem A at the end of~\cite{Lee:fredholm}, compare~\cite{GL} for  the Poincar\'e ball):

 \begin{proposition}
   \label{p10III17.1}
   Suppose that $\zgriem$ is non-degenerate and that
   $\zP h=0$, for a tensor field $h$ satisfying
\beal{10III17.41}
 &
 |h|_\zgriem = o(1)
%  \,.
 &
\eeal{}
as the conformal boundary is approached. Then $h \equiv 0$.
 \end{proposition}

Our solutions will be perturbations of a space-time $(\mcM,\zglorentz) $ with a static metric $\zglorentz$ solving the vacuum Einstein equations with a negative cosmological constant. By definition of staticity, near every point in $(\mcM,\zglorentz)$ at which the Killing vector is timelike there exist coordinates in which the metric  takes the form \eq{8III17.1}-\eq{8III17.2} with $\theta \equiv 0$,
\beal{14I17.6}
 &\zglorentz  = -\mathring V^2 dt ^2 + \underbrace{\mathring g_{ij}dx^i dx^j}_{=:\mathring g}\, ,
 \quad
 \partial_t \mathring V =  \partial_t \mathring g=0\, .
 &
\eea
The solutions we are about to construct will be defined in the domain of outer communications, where the representation \eq{14I17.6} is in fact global.

In this work we will consider two cases:

\begin{enumerate}
\item[H1.]
 \label{p6III17.1}
 $\mathring V$ is strictly positive, $\mcM$ is diffeomorphic to $\R\times M$, where the coordinate $t$ along the $\R$ factor labels the static slices of $\zglorentz$ in $\mcM$. We set
 $$
  \Mriem= S^1\times M
   \,,
  $$
 % ,
 thus time translations in $\mcM$ become rotations of the $S^1$-factor of $\Mriem$.

 In this case our analysis below provides an alternative proof of the results in~\cite{ChDelayStationary,ChDelayKlinger}.
\item[H2.]
 \label{p6III17.3}
  We allow $\zglorentz$ to describe a static vacuum black-hole metric with a Killing horizon with non-zero surface gravity and with global structure similar to that of the domain of outer communications in the Schwarzschild-anti de Sitter (S-AdS) black holes. More precisely, we assume that the Lorentzian manifold $\mcM$ takes the form
%   \ptc{notation for the manifold $\ourN$ changed, by a macro, to avoid possible confusions with %$\Nman$ in Section~\ref{sA7IV17.1}}
%
$$
   \mcM=\R\times [R_0,\infty)\times \ourN  \,,
$$
for some $R_0>0$,  where $\ourN $ is a compact $(n-1)$-dimensional boundaryless manifold.   We require that $\R\times \{R_0\} \times \ourN $  coincides with the zero-level set of $\zV$ which, in a suitable extension of $\mcM$, becomes an event horizon with non-zero surface gravity. The coordinate $t$ along the $\R$-factor labels the static slices of $\mcM$. We further assume that after a ``Wick rotation'', where $dt^2$ is replaced by $-dt^2$, the resulting Riemannian metric
\bel{14I17.8}
 \zgriem:= \zV^2 dt^2 + \mathring g
 \,.
\ee
  extends to a smooth metric  on
     $$
      \Mriem =  \R^2 \times \ourN
       \,,
     $$
 with the action of the flow of the vector field
 \bel{10III17.7}
  X:=\partial_t
 \ee
 %$
 being rotations of the $\R^2$ factor.
% We further assume that the factors $\R^2$ and $\ourN $ are orthogonal for $\zgriem$.
%  \ptcr{last one probably not needed}

  In this case our analysis generalises the results in~\cite{ChDelayStationary,ChDelayKlinger} to black-hole solutions.

%
%
%\item
% \label{p6III17.3}
%  \ptcr{all blue is old, to be thought over, probably to discard, and to be considerably rewritten if kept}
%  \blue{We allow $\zglorentz$ to describe a black-hole metric with a non-degenerate Killing horizon. To be specific, let $(M,g)$ be a manifold with a boundary with exactly two components.\ptc{allow more?}
%     The first, denoted by $\partial_\infty M$, is a conformal boundary near which $g$ is asymptotically hyperbolic, with $\zV^{-2}\zg$ extending smoothly through $\partial_\infty M$. The remaining component  of $\partial M$ will be denoted by  $N$. We require that it coincides with the zero-level set of $\zV$, and that $N$ corresponds to the bifurcation surface of a bifurcate Killing horizon for the Killing vector $\partial_t$ (for definitions, see e.g.~\cite{CCH}. On the manifold
%     %
%     $$
%      \Mriem = (S^1 \times M)/\sim
%       \,,
%     $$
%     %
%     where the equivalence relation $\sim$ identifies points $(t_1,p)$ with $(t_2,p)$ when $p\in N$, we define  the metric
%%
%\bel{14I17.8old}
% \zgriem = \zV^2 dt^2 + \mathring g
% \,.
%\ee
%%
%It is well known that a suitable normalisation of the Killing vector $\partial_t$ can be chosen so that $(\Mriem,\zgriem)$ is smooth, with the action of the flow of $\partial_t$ being rotations in the two-dimensional bundle of normals to $N$.
% }
\end{enumerate}

An example of  H2 above is given by the $(n+1)$-dimensional Schwarzschild-anti de Sitter metrics with non-vanishing surface gravity, where $\ourN $ is the $(n-1)$-dimensional sphere $S^{n-1}$ and  $\Mriem = \R^2 \times S^{n-1}$.
%We show in Appendix~\ref{sA7IV17.1} that four-dimensional Schwarzschild-anti de Sitter black holes
%\red{
%with non-vanishing surface gravity and with mass parameter satisfying
%%
%\bel{9VII17.1c}
% 0<\mhere <\ell \ \mbox{ and }
%\ \mhere
% \ne
% 2 \, \ell\, 3^{-\frac 32}
%\ee
%%
%%
%are non-degenerate (the reader is warned that, for ease of comparison of the calculations in our Appendix with~\cite{IshibashiKodamaStability}, $n_{\mathrm{Appendix}} = n_{\mathrm{remainder\ of\ the\ paper}}-1$).}
More generally, the $(n+1)$-dimensional Birmingham metrics \cite{Birmingham}, where $\ourN $ is an $(n-1)$-dimensional Einstein manifold, with non-extreme horizons are of this form.
%, cf.\ Appendix~\ref{sA7IV17.1}.%
% When $\ourN $ is  negatively curved, the resulting Riemannian manifold $(\Mriem,\zgriem)$ is non-degenerate for a whole range of values of the mass parameter $\mhere $~\cite[Proposition~D.2]{ACD2}.\pk{remove this sentence? obsolete and now mentioned in introduction}

\section{The construction}
 \label{s8III17.1}

To avoid a discussion of the technicalities associated with the matter fields, we will start by describing in some detail the construction of the vacuum solutions.
Note, however, that the argument is essentially the same in both cases, once the isomorphisms needed to handle matter fields have been established. The key difference is in the boundary conditions: the vacuum stationary solutions are determined by their asymptotic data at the conformal boundary, and might have a non-standard conformal infinity when these data are \emph{not} the usual AdS ones.%
\footnote{Note that some non-trivial asymptotic data are compatible with the usual locally conformally flat structure of the conformal boundary. An example is provided by the Demia\'nski-Carter  ``Kerr anti-de Sitter'' solutions, see \cite[Appendix~B]{HT}.}
On the other hand, our non-vacuum solutions are determined by both the asymptotic data for matter fields and for the metric, which allows existence of nontrivial solutions with the manifestly standard AdS conformal structure at timelike infinity.

 \subsection{Vacuum solutions}

We denote by $\rho$  a coordinate near $\partial M$ which vanishes at $\partial M$, and by  $C^{\ourell ,\alpha}(\partial M,{\mathcal
 T}_1)$ the space of one-forms on $\partial M$ of $C^{\ourell ,\alpha}$-differentiability class.

We have the following:

\begin{theorem}\label{T8III17.1}
 Let $n=\dim M\ge 3$,
   $k\in\N\setminus\{0\}$, $\alpha\in(0,1)$, and
 consider a static Lorentzian real-valued Einstein metric $\zglorentz$ of the form
 \eq{8III17.1}-\eq{8III17.2} as described in Section~\ref{s8III17.0}, such that the associated Riemannian
 metric $\zgriem$ is $C^2$ compactifiable and non-degenerate, with
 smooth conformal infinity.  We further assume that the hypotheses H1 or H2 of Section~\ref{s8III17.0} hold. For all $a\in\R$ with $|a|$ small enough and every smooth real-valued $\hat{\theta} \in C^{k+2,\alpha}(\partial M,{\mathcal
 T}_1)$  there exists a unique, modulo diffeomorphisms which are the
 identity at the boundary, nearby stationary Lorentzian real-valued vacuum metric of the form
 \eq{8III17.1}-\eq{8III17.2} such that, in local coordinates near the
 conformal boundary $\partial M$,
\bel{8III17.3}
 V-\mathring
 V=O(\rho)\,,\quad
  \theta_i=a \hat \theta_i +O(\rho)
 \,,\quad
 g_{ij}-\mathring g_{ij} =O(1)\,.
\ee
\end{theorem}

The Lorentzian solutions with $V>0$ (case H1) are globally stationary, in the sense that they have a globally timelike Killing vector. We show in Section~\ref{s10III17.1} below that the Lorentzian solutions with $V$ vanishing (case H2)  describe smooth black holes.

\begin{remark}
  \label{r8III17.1}
  Large families of static vacuum metrics $\glorentz$ satisfying the conditions of the theorem have been constructed in~\cite{ACD2,ACD}. In particular if $\glorentz$ is non-degenerate, then the nearby metrics as constructed in~\cite{ACD2,ACD} also are.
\end{remark}

\begin{remark}
  \label{r8III17.2}
   The decay rates in \eq{8III17.3} have to be compared with the
leading order behavior $\rho^{-2}$ both for ${\mathring V}^2$ and
${\mathring g}_{ij}$  in local coordinates near the conformal boundary. A precise version of \eq{8III17.3} in terms of
weighted function spaces  reads (our notation for function spaces follows~\cite{Lee:fredholm})
\beal{8III17.3a}
 & (V-\mathring V)\in
C_{1}^{k+2,\alpha}(\Mriem)\,,\;\;(g-\mathring g)\in
C_{2}^{k+2,\alpha}(\Mriem,{\mathcal S}_2)\,,
 & \\ &
 \quad
\theta-a \hat \theta \in C_{2}^{k+2,\alpha}(\Mriem,
{\mathcal T}_1)\,,
 &
\eeal{8III17.3b}
and the norms of the differences above are small in those spaces. If the boundary data are smooth, then the solution has a complete polyhomogeneous expansion at the conformal boundary.
\end{remark}

{\sc \noindent Proof:} We start by solving on $\Mriem$  the ``harmonically-reduced Riemannian Einstein equations'',
\bel{10III17.8}
  \Rriem^H_{\alpha\beta} +n \griem_{\alpha\beta} =0
  \,,
\ee
for a complex-valued tensor-field $\griem$,
 with the asymptotic conditions
\bel{8III17.5}
 V-\mathring
 V=O(\rho)\,,\quad \theta_k= i a \hat \theta_k +O(\rho) ,\quad
 g_{k\ourell }-\mathring g_{k\ourell } =O(1)\,.
\ee
Here we have extended $\hat \theta$ from $  \partial M$  to $S^1\times \partial M$ by imposing invariance under rotations of the $S^1$ factor.

The  existence of a solution, for all $a\in \C$ with $|a|$ small enough, follows by rewriting the equations  as in \eq{14I17.2}-\eq{14I17.3} (with $({\Phi}^A)= (\griem_{\mu\nu})$), and applying the implicit function theorem. This can be done because of our hypothesis of non-degeneracy of $\zgriem$; see~\cite{ACD2,ChDelayStationary} for the analytical details. In particular \eqref{8III17.3a}-\eqref{8III17.3b} hold.

The implicit function theorem guarantees that the solutions sufficiently close to $\zgriem$ with the asymptotics  \eq{8III17.5} are uniquely determined by $a\hat \theta$.  We  denote by
$\griem(a)$ this solution.

The usual argument, spelled-out in detail e.g.\ in~\cite[Section~4]{ChDelayKlinger},
 applies to show that $\lambda^\mu \equiv 0$, so that:

\begin{lemma}
  \label{L10III17.4}
The complex metrics $\griem(a)$ solve the Riemannian Einstein equations.
\qed
\end{lemma}

We continue by showing that:

\begin{lemma}
  \label{L10III17.3}
The complex metrics $\griem(a)$ are invariant under rotations of the $S^1$ factor of $\Mriem$ in the case H1, or of the $\R^2$ factor in the case H2.
\end{lemma}

\proof
Let us denote by $P(a)$ the operator obtained by  linearising  \eq{10III17.8} at $\griem(a)$; compare \eq{10III17.5}. The Lie derivative of \eq{10III17.8} with respect to $X$ gives
\bel{8III17.12}
 P(a) \mcL_X \griem(a) =0
 \,,
\ee
where $\mcL_X$ is the Lie-derivative with respect to the vector field $X$ generating rotations of the $S^1$ factor of $\Mriem$ in the case H1, or of the $\R^2$ factor in the case H2; we have also used the fact that $\mcL_X \zgriem = 0$. It follows from \eq{8III17.3a}-\eq{8III17.3b} and polyhomogeneity of the solutions that
\beal{8III17.3a+}
 & \mcL_X V  \in
C_{1}^{k+1,\alpha}(\Mriem)\,,\;\;  \mcL_X g  \in
C_{2}^{k+1,\alpha}(\Mriem,{\mathcal S}_2)\,,
 & \\ &
 \quad
 \mcL_X  \theta \in C_{2}^{k+1,\alpha}(\Mriem,
{\mathcal T}_1)\,.
 &
\eeal{8III17.3b+}
This, together with Proposition~\ref{p10III17.1}, implies $\mcL_X \griem \equiv 0$, as desired.
\qedskip

Denoting by $t$ the usual angular coordinate on the $\R^2$ factor (H2 case), or the parameter along $S^1$ (H1 case), we can thus write the metrics $\griem(a)$ in  coordinates adapted to the flow of $X$ in the form
\beal{20XI16.23+}
 & \griem(a):=   V(a)^2(dt+ a i \theta(a)_k dx^k )^2 +g(a)_{ij}dx^idx^j \, .
 &
\eea

\begin{lemma}
  \label{L10III17.2}
In  coordinates as in \eq{20XI16.23+},  the functions $V(a)$, $g_{ij}(a)$ and  $\theta(a)_i$ are even functions of $a$.
\end{lemma}

\proof
Let $\psi:\Mriem\to\Mriem$ denote the map which, in the coordinates of \eq{20XI16.23+} changes $t$ to its negative, leaving the remaining coordinates unchanged. Then $\psi$ is a smooth isometry of $(\Mriem,\zgriem)$. The metric $
\psi^*\griem(a)$ satisfies the same equation, with same asymptotic data, as   $\griem(-a)$, and is close to $\zgriem$ for $|a|$ sufficiently small, so that  uniqueness gives
\bel{10III2017.22}
 \psi^*\griem(a)  = \griem(-a)
  \,,
\ee
which implies the claim.
\qedskip

\begin{lemma}
  \label{L10III17.1}
  The metrics $\griem(a)$ are holomorphic functions of $a$.
\end{lemma}

\proof
It is standard to show that the metrics $\griem(a)$ are continuously differentiable functions of $a$. Differentiating  \eq{10III17.8} with respect to $\overline a$ gives
$$
 P(a) \frac{\partial \griem(a)}{\partial \overline a} = 0
 \,,
$$
{where $\partial/\partial \overline a$ is the usual complex-derivative operator with respect to the complex conjugate $\overline a$ of $a$.
The  vanishing of the asymptotic data for $\frac{\partial \griem(a)}{\partial \overline a}$ gives  $\frac{\partial \griem(a)}{\partial \overline a} \equiv 0$.
\qedskip

Now, if $a \in i \R$, we can repeat the above construction in the space of real-valued Riemannian metrics. Uniqueness implies  then that the corresponding metrics $i\R\ni a\to\griem(a)$ are real-valued. Hence all the  coefficients $\griem(x^\alpha)_{\mu\nu k }$ in the convergent Taylor expansions
\bel{10III17.11}
 \griem(a,x^\alpha)_{\mu\nu} = \sum_{k\in \N} \griem(x^\alpha)_{\mu\nu k } (ia)^k
\ee
are real.  Lemma~\ref{L10III17.2} implies  that $V(a)$, the $g_{ij}(a)$'s, and the $\theta(a)_i$'s are real for real $a$. It follows that for real $a$ the real-valued  Lorentzian metrics
\bel{10III17.13}
 \glorentz(a):=   -V(a)^2(dt+ a   \theta(a)_k dx^k )^2 +g(a)_{ij}dx^idx^j
\ee
satisfy all our claims.
\qedskip

 \subsection{Matter fields}
  \label{ss10III17.1}

We now seek solutions to the Einstein-Yang-Mills-Higgs-Maxwell-dilaton-scalar fields-Chern-Simons equations defined by the action
 \bel{21XI16.2}
 S=\int d^{n+1}x \frac{\sqrt{-\det{}\glorentz }}{16\pi G}\left[R(\glorentz)-2\Lambda-W({\Phi})|F|^2-\frac{1}{2}(\nabla{\Phi})^2-{\mycal V}({\Phi})\right] +S_{\mathrm{CS}}\,.
\ee
Here $R(\glorentz)$ is the Ricci scalar of the metric $\glorentz$, $W$ and ${\mycal V}$ are smooth functions, $|F|$ is the gauge-invariant norm of a possibly non-Abelian Yang-Mills field, ${\Phi}$ is allowed to be a section of a bundle associated to the possibly non-Abelian gauge-group, with $\nabla {\Phi}$ depending if desired upon the Yang-Mills gauge potential.  Finally, in even space dimension $n$,  $S_{\mathrm{CS}}$ is the Chern-Simons action which, in the Abelian case, takes the form:
\bel{8XII16.2}
 S_{\mathrm{CS}} = \left\{
                     \begin{array}{ll}
                       0, & \hbox{$n$ is odd;} \\
 \displaystyle
                       \frac{\lambda }{16\pi G}\int A\wedge \underbrace{F\wedge \cdots \wedge F}_{k \text{ times}}, & \hbox{$n=2k$,}
                     \end{array}
                   \right.
\ee
for a constant $\lambda \in \R$.
In the general (non-Abelian) case $S_{\mathrm{CS}}$ is given by~\cite[Equation~(3.5)]{ChernSimons}
\bel{20III17.1}
S_{\mathrm{CS}}=\frac{\lambda}{16\pi G} \int \Tr\left(\sum_{i=0}^{k}C_{k,i} A\wedge [A,  A]^i \wedge F^{k-i}\right)\,,
\ee
when $n=2k$ and
\bel{20III17.2}
C_{k,i}=\frac{(-1)^i(k+1)!k!}{2^i(k+1+i)!(k-i)!}\,.
\ee

We obtain:
	\begin{theorem}\label{T14III17.1}
 Let $n=\dim M\ge 3$,
   $k\in\N\setminus\{0\}$, $\alpha\in(0,1)$, and
 consider a static Lorentzian real-valued Einstein metric $\zglorentz$ of the form
 \eq{8III17.1}-\eq{8III17.2} as described in Section~\ref{s8III17.0}, such that the associated Riemannian
 metric $\zgriem$ is $C^2$ compactifiable and non-degenerate, with
 smooth conformal infinity, has no harmonic one-forms which are in $L^2$, with  $\mcV''(0)$ which
is not an $L^2$-eigenvalue of the operator $\Delta_{\mathring\griem}$.
%(cf.\ Remark~\ref{R31III17.1} below).
   We further assume that the hypotheses H1 or H2 of Section~\ref{s8III17.0} hold and that
 \bel{8XII16.+1}
 W(0)=1
 \,,
  \quad
  \mcV(0)=0
   =
  \mcV'(0)
  \,,
   \quad
   \mcV''(0)> -n^2/4
  \,.
\ee
  For all $a\in\R$ with $|a|$ small enough, every smooth real-valued $\hat{\theta} \in C^{k+2,\alpha}(\partial M,{\mathcal T}_1)$ and $\hat U \in C^{k+2,\alpha}(\partial M)$ and

\begin{enumerate}
  \item
$
\mcV''(0) < 0
$
with $\hat A\in C^{k+2,\alpha}(\partial M,{\mathcal T}_1)$, and  $\hat {\Phi} \in \rho^{\sigma_-} C^{k+2,\alpha}(\partial M)$ (where $\sigma_-=n/2-\sqrt{n^2/4+\mcV''(0)}$) which are sufficiently small smooth fields on $\partial  M$, or

  \item
$\hat {\Phi} \equiv 0$, and $\hat A \in C^{k+2,\alpha}(\partial M,{\mathcal T}_1)$ which is a sufficiently small smooth field  on $\partial M$,
\end{enumerate}

\noindent
there exists a unique, modulo diffeomorphisms which are the
 identity at the boundary, nearby stationary Lorentzian solution of the Einstein-Maxwell-dilaton-scalar fields-Chern-Simons equations, or of the Yang-Mills-Higgs-Chern-Simons-dilaton equations  with a trivial principal bundle, so that, in local coordinates near $\partial M$, we have
\bel{27XII16.1}\begin{split}
 g\to_{\rho \to 0} \mathring g
 \,, \
 V \to_{\rho \to 0} \mathring V
  \,, \ \theta \to_{\rho \to 0} a \hat \theta \,, \\
U\to_{\rho \to 0} a \hat U
 \,, \ A \to_{\rho \to 0} \hat A_a dx^a
 \,, \ {\Phi} \to_{\rho \to 0} \hat {\Phi}
 \end{split}
\ee
with all convergences in $\mathring g$-norm. The hypothesis of non-existence of harmonic $L^2$-one-forms is not needed if $\hat A\equiv 0 \equiv \hat U$, in which case the Maxwell field or the Yang-Mills field are identically zero.
\end{theorem}

\begin{remark}
  \label{R31I17.2}
  The remarks in~\cite[Section~7]{ChDelayKlinger} concerning the energy and the asymptotics of the solutions remain valid word-for-word in the current setting.
\end{remark}

\begin{remark}
 \label{R24III17.1}
For the convenience of the reader we repeat  here the comments from~\cite{ChDelayKlinger} concerning the kernel conditions in the theorem.

First, it is shown in~\cite[Appendix~C]{ChDelayKlinger} that the condition of non-existence of $L^2$-harmonic forms is satisfied near anti-de Sitter space-time in any case.

Next, it has been shown by Lee~\cite[Theorem A]{lee:spectrum} that there are no $L^2$-eigenvalues of $\Delta_{\mathring \griem}$
% Assume that $f$ is in $L^2(d\mu_{\mathring g})$ and in the kernel of $T_1+\omega^2$.
%Then $\sigma=V^{-1/2}f$ is in  $L^2(Vd\mu_{\mathring g})$ and in the kernel of $\mathcal %T_1+\omega^2$.
%This implies that $\sigma$ is an $L^2(d\mu_{\mathring \griem})$-eigenfunction of $\Delta_{\mathring\griem}$ for the eigenvalue $-\omega^2$. In particular if there are no such eigenvalues, the hypotheses of Corollary~\ref{isofunctioncoro} are verified.
%For instance, from~\cite{lee:spectrum}, Theorem A,
%this is always the case
when the Yamabe invariant of the conformal infinity is
positive, in particular near anti-de Sitter and Schwarzschild anti-de Sitter space-time.  Furthermore, and quite generally,  $\mcV''(0)=0$ is never an eigenvalue by the maximum principle.  Finally, again quite generally, the $L^2$ spectrum of $-\Delta_{\griem}$ for asymptotically hyperbolic manifolds
is $[n^2/4,+\infty[$ together with possibly a finite set of eigenvalues, with finite multiplicity, between $0$ and $n^2/4$~\cite{Guillarmou} (compare~\cite{MazzeoMelrose}),
so our non-eigenvalue  condition is true except for at most a finite number of values of $\mcV''(0)\in (-n^2/4,0)$ for all asymptotic geometries.
%Finally for a concrete test of this condition, we recall that if $-\omega^2$ is an $L^2$ eigenvalue of $T_1$,  then the corresponding eigenfunction is polyhomogeneous and exactly of order $O(\rho^{\delta^+})$, where
%%
%$$
%\delta^+=\frac n 2+\sqrt{\left(\frac{n}2\right)^2-\omega^2}.
%$$
%
\qed
\end{remark}

   \begin{proof}
   This follows directly from the arguments of~\cite{ChDelayKlinger}: The indicial exponents of the relevant equations remain unchanged, as terms containing $\theta$, which are here multiplied by $ia$, are of lower order in $\rho$. Note that the solutions obtained below using the implicit function theorem might a priori depend upon the ``periodic time coordinate'' $t$, but this is irrelevant for the calculation of the indicial exponents.

 We start by sketching the argument in the case of a single real-valued  scalar field ${\Phi}$, which satisfies the equation
  \[
   \nabla_\alpha\nabla^\alpha {\Phi}-\mcV'({\Phi})=0\,.
   \]
    Its indicial exponents are $\sigma_\pm=n/2\pm\sqrt{n^2/4+\mcV''(0)}$, unchanged from those in~\cite{ChDelayKlinger}, as $\griem^{tt}=-V^{-2}=O(\rho^2)$ so that terms arising from $t$ derivatives are of lower order.
We assume $-n^2/4<\mcV''(0)<0$ so that the solutions show the desired asymptotics. By~\cite[Theorem~D.1]{ChDelayKlinger}, using the assumption that $\mcV''(0)$ is not an $L^2$ eigenvalue of $\Delta_{\mathring \griem}$, it follows that the linearisation $(\Delta_{\mathring \griem}-\mcV''(0))$ is an isomorphism from $C_{\sigma_-+s}^{k+2,\alpha}$ to $C^{k,\alpha}_{\sigma_-+s}$ for small $s>0$.

Then $\mcL_X {\Phi}\equiv 0$ by the same argument as in the proof of Lemma \ref{L10III17.3} above: Applying $\mcL_X$ to the equations and using $\mcL_X \mathring\griem=0$ gives
    \bel{21III17.10}
    \left\{\begin{array}{l}
    P(a)\mcL_X\griem=\mcL_X \big( q_1[V - \mathring V, g - \mathring g, \theta, {\Phi}]
    \big) \,,\\
   (\Delta_{\mathring \griem}-\mcV''(0))\mcL_X {\Phi}=q_2[V, g, \theta, {\Phi}]\,.
   \end{array}\right.
   \ee
Here $q_1$ is a linear combination of the energy-momentum tensor of the scalar field and its trace times the metric, and is at least quadratic in its arguments and their derivatives, so that $\mcL_X q_1$ is a linear first-order differential operator in $(\mcL_X\griem,\mcL_X {\Phi})$. Furthermore, each term in $q_2$ is linear in ${\Phi}$ or its derivatives and contains an $\mcL_X$ derivative of one of the arguments. ${\Phi}$ behaves asymptotically as in~\cite{ChDelayKlinger}, i.e. ${\Phi}=\rho^{\sigma_-}\hat{\Phi}+o(\rho^{\sigma_-})$ and, using $\mcL_X \hat{\Phi}=0$, we have $\mcL_X  {\Phi} = o(\rho^{\sigma_-})$. As  the coefficients of $\mcL_X {\Phi}$ and $\mcL_X \griem$, and of their first derivatives,  on the right-hand side of \eqref{21III17.10} are small in the relevant spaces (e.g. the coefficients of the $\mcL_X {\Phi}$ terms on the right-hand side of the second equation are small in $C^{k,\alpha}_0$), we can use Proposition \ref{p10III17.1} and the isomorphism properties of $(\Delta_{\mathring \griem}-\mcV''(0))$ described above
to conclude that $\mcL_X \griem\equiv 0$ and $\mcL_X {\Phi} \equiv 0$.

 Next, we show that $V, g, \theta, {\Phi}$ are even functions of $a$: $\psi^*(\griem(a), {\Phi}(a))$ satisfy the same equations as $(\griem(-a), {\Phi}(-a))$, with identical asymptotic data (since $\hat {\Phi}$ is independent of  $a$, the only relevant terms are the asymptotic data for  $\griem(a)^{0j}=- i a g(a)^{jk}\theta(a)_k$ which are unchanged under $(t,a)\mapsto (-t,-a)$)
 and by uniqueness we have ${\Phi}(a)=\psi^*{\Phi}(a)={\Phi}(-a)$, similarly for $V$, $g$, $\theta$.

Therefore ${\Phi}(a)$, $V(a)$, $g_{ij}(a)$, $\theta_i(a)$ are real for real $a$ by the same argument as for $V(a)$, $g_{ij}(a)$, $\theta_i(a)$  above.

Rather similar considerations apply for the whole system of Einstein-Maxwell-dilaton-Yang-Mills-Higgs-Chern-Simons-scalar fields equations: The matter equations arising from the action \eqref{21XI16.2} are
\begin{equation}\label{matterequations}
\left\{\begin{array}{l}
\frac{1}{V\sqrt{\det g}}\partial_\mu (V \sqrt{\det g} W F^{\mu\nu})+B_{\mathrm{CS}}^\nu=0\,,\\
\frac{1}{V \sqrt{\det g}}\partial_\mu (V \sqrt{\det g} \griem^{\mu\nu} \partial_\nu {\Phi}) - W'({\Phi}) |F|^2 - {\mycal V}'({\Phi})=0\,,
\end{array}\right.
\end{equation}
where%
\footnote{One can check by a direct time-and-space decomposition of the equations that the ``Wick rotation'' $dt\to -idt$, $\partial_t\to i \partial_t$, is consistent with the Chern-Simons terms in the equations by defining $\epsilon^{\alpha_1 \ldots \alpha_d}$ as $(-\det \glorentz)^{-1/2}\mathring\epsilon^{\alpha_1 \ldots \alpha_d}$, where $\mathring \epsilon^{\alpha_1 \ldots \alpha_k}$ is totally antisymmetric with values in $\{0,\pm 1\}$,  with the cut in the definition of $\sqrt{z}$,  $z\in\C$, lying e.g. on the positive imaginary axis, so that  $\sqrt{z^2}= z$ both near $z=-i$ and near $z=1$.}
\bel{8XII16.3}
 \displaystyle
B_{\mathrm{CS}}^\nu = \left\{
                     \begin{array}{ll}
                       0, & \hbox{$n$ is odd;}
 \\
 \displaystyle
                  -  \frac{ \lambda }{2^{k+2}}
\epsilon^{\nu\alpha_1\beta_1\cdots\alpha_k\beta_k} F_{\alpha_1\beta_1} \cdots F_{\alpha_k\beta_k}, & \hbox{$n=2k$.}
                     \end{array}
                   \right.
\ee

 After replacing $dt$ by $-i dt$, the asymptotic data for the Riemannian solution, say $\hat F(a)$, take the form $\hat F(a)=d(- i a \hat U  dt+\hat A_i  dx^i)$. They are  clearly invariant under
\bel{31III17.5}
 \mbox{$t\mapsto -t$ and $a \mapsto -a$.}
\ee

The only other asymptotic data that are possibly affected by \eq{31III17.5} are those associated with the inverse metric components $\griem(a)^{0j}$. These  change sign under each of $a\mapsto -a$ and $t\mapsto -t$. It follows
$$
 \psi^*(F(-a),\griem(-a),{\Phi}(-a))= (F(a),\griem(a),{\Phi}(a))
  \,,
$$
which again implies that $(U(a),A(a) ,{\Phi}(a),V(a),\theta(a),g(a))$ are even functions of $a$. As before, analyticity holds and we conclude that   all these   fields  are  real for real $a$.

Note that the implicit function theorem in the Riemannian regime produces essentially complex electric fields   for real $a$ and non-zero $\hat U$'s, which will however be mapped to real ones when one returns to the Lorentzian setting.
   \end{proof}

\subsection{$f(R)$ theories}
 \label{ss7IV17.1}
Our method allows the construction of black-hole solutions to specific $f(R)$ theories: As described in e.g.~\cite[Section~2.3]{DeFelice2010} these can be reduced to the Einstein-scalar field equations with a specific potential $\mcV({\Phi})$ by a conformal transformation, if the function $f$ fulfills certain conditions. These conditions are satisfied simultaneously with our assumptions on $\mcV$ (in Theorem \ref{T14III17.1}) if
\begin{equation}
f'>0\,,\quad f''\neq 0
 \,,\quad  f'^{-1}(1)<0
  \,,\quad f(f'^{-1}(1))=f'^{-1}(1)/2
 \,.
\end{equation}
(this is shown in detail in~\cite[Section~5.5]{ChDelayKlinger}). An example of a function $f$ which fulfills these conditions is
\begin{equation}
 f(\tilde{R})=d \tilde{R}+c \tilde{R}^{\alpha+1} + e
  \,,
\end{equation}
where $\tilde R$ is the Ricci scalar in the $f(\tilde R)$ theory (i.e. before the conformal transformation) and $d<1$, $c<0$, $\alpha=1,3,5,\dots$, and
\[
e=\frac{\alpha(1-2d)-1}{2(\alpha+1)} \sqrt[\alpha]{\frac{1-d}{c(\alpha+1)}}\,,
\]
are constants.

\subsection{Time-periodic scalar fields}
 \label{ss3IVl17.1}
Similarly to~\cite[Section 6.1]{ChDelayKlinger} we can use the method there to construct solutions with a time-periodic complex scalar field ${\Phi}(t,x)=e^{i \omega t}\psi(x)$ where $\omega\in \R$ is a constant and $\psi(x)$ is allowed to be complex. We assume that
\bel{12I17.1}
 \mbox{$\mcV ({\Phi})=\GmcV ({\Phi} \bar{\Phi})$ and $W({\Phi})=\GW ({\Phi}\bar{\Phi})$}
\ee
for some differentiable functions $\GmcV$ and $\GW$, and replace the term $(\nabla {\Phi})^2 $ in the action by $\nabla^\alpha \overline{{\Phi}}\nabla_\alpha {\Phi} $, where $\overline{{\Phi}}$ is the complex conjugate of ${\Phi}$.

The Lorentzian ${\Phi}$ equation for a complex scalar field ${\Phi}=e^{i \omega t}\psi$ takes the form
\bel{4IV17.7}
\Delta_\glorentz (e^{i\omega t}\psi)-\GW'(\psi\bar\psi)e^{i\omega t}\psi |F|^2-\GmcV'(\psi\bar\psi)e^{i\omega t}\psi=0
    \,.
\ee
This leads to the following associated Riemannian equation
\beal{4IV17.8}
 &
 \Delta_\griem (e^{a\omega t}\psi)-\GW'(\psi\bar\psi)e^{a\omega t}\psi |F|^2-\GmcV'(\psi\bar\psi)e^{a\omega t}\psi=0
 &
\\
&
 \Longleftrightarrow
 &
 \nn
\\
 &\begin{split}
 0=&\Delta_\griem \psi + 2   a\omega \griem(D t, D \psi) +|dt|^2_{\griem} a^2\omega^2 \psi
  +
  a \omega \psi \Delta_\griem t
   \\
 &
  -\GW'(\psi\bar\psi) \psi |F|^2-\GmcV'(\psi\bar\psi) \psi \psi
 %
% +a\omega \psi \partial_t(V^{-2}-\theta_j\theta^j a^2)+i a^2 \omega \psi \theta^j{}_{,j}\\
% &+(V^{-2}-\theta_j\theta^j a^2) a\omega \psi \partial_t\log(V\sqrt{\det g})+i a^2 \omega\psi \theta^j\partial_j\log(V\sqrt{\det g})
\,,
 \end{split}
 &
 \nn
\eea
where the crucial difference to a naive replacement $t\mapsto -i t$ (and therefore ${\Phi}\mapsto e^{\omega t}\psi$) is that the argument of $\GW'$ and $\GmcV'$ is $\psi\bar\psi$ instead of ${\Phi}\bar{\Phi}=e^{2\omega t}\psi\bar\psi$. The equations \eqref{4IV17.7} and \eqref{4IV17.8}, together with the respective Lorentzian and Riemannian equations for the other variables, are equivalent: The bijection
\bel{4IV17.10}
(V, \theta, g, U, A, \omega, \psi)\mapsto (i V,  i \theta, g, - i  U, A, - i \omega, \psi)
\ee
maps Lorentzian solutions to Riemannian ones.

As such, the first equation \eq{4IV17.8} does not make sense for periodic $t$'s, but the second does. Note, however, that $Dt$ and
$|dt|^2$ are singular at an axis of rotation of   $\partial_t $, if there is one. This forces us to restrict ourselves to strictly stationary configurations, without black holes, when $\omega\ne 0$. As a consequence, in this section we merely reproduce the results already proved in~\cite{ChDelayKlinger} for rotating complex fields, albeit by a somewhat simpler argument.

Applying $\mcL_X$ to the second equation in \eqref{4IV17.8} gives
%
%\bel{4IV17.9}\begin{split}
%0=&\omega\overbrace{\left(\Delta_\griem (e^{\omega t}\psi)-\GW'(\psi\bar\psi){\Phi} |F|^2-\GmcV'(\psi\bar\psi){\Phi}\right)}^{=0}\\
%&+\Delta_\griem(e^{\omega t}\mcL_X \psi)
%+O(\psi e^{\omega t}\mcL_X \griem)\\
%&-e^{\omega t}\mcL_X\left(\GW'(\psi\bar\psi)\psi |F|^2+\GmcV'(\psi\bar\psi)\psi\right)\,.
%\end{split}
%\ee
%
%
\[\begin{split}
0=&\Delta_\griem(\mcL_X \psi) +2\omega \griem(Dt,D\mcL_X \psi)\\
&-\mcL_X\left(\GmcV'(\psi\bar\psi)\psi+\GW'(\psi\bar\psi)\psi |F|^2\right)\\
&+O(\psi\mcL_X\griem) +O(D\psi \mcL_X \griem)+ O(DD\psi \mcL_X \griem)+O(\rho\mcL_X\psi)\,,
\end{split}\]
and therefore
\bel{20IV17.1}
 (\Delta_{\mathring \griem}-\GmcV'(0))\left(\mcL_X \psi \right)=q_3[\griem, \psi, U, A]\,,
\ee
where each term in $q_3$ is at least linear in its arguments or their derivatives and contains an $\mcL_X$ derivative of $\griem$, $\psi$, $U$, or $A$.

We can now argue as before to obtain $\mcL_X (\griem, \psi, U, A)\equiv 0$ if the asymptotic data are invariant under $\mcL_X$.

The equation for $\psi$ is then
\bel{3IV17.5}
\begin{split}
&V^{-1}D_i (Vg^{ij}\partial_j \psi)
- \big( \GW'(\psi\bar\psi)|F|^2
 + \GmcV'(\psi\bar\psi)\big) \psi\\
&\qquad+(V^{-2}-a^2\theta_k\theta^k) a^2 \omega^2 \psi+ i a^2\omega (\theta^j \partial_j\psi+V^{-1}D_j(V\theta^j \psi))=0\,.
 \end{split}
\ee

All terms in this equation are well defined and, by the results of~\cite{ChDelayKlinger}, we obtain a solution of the complete system of equations. As $\psi$ is independent of $t$, no difficulties associated with the periodicity of the $t$ coordinate arise. After transforming back via the inverse of \eqref{4IV17.10} we obtain a time-periodic solution ${\Phi}(t,x)=e^{i\omega t}\psi(x)$ to the original equations.

\section{Geometry of the solutions}
  \label{s10III17.1}

We wish to show that the solutions constructed above with topology $\Mriem =\R^2 \times \ourN $ correspond to smooth black holes on the Lorentzian side. (In fact, the Lorentzian metric will be one-sided-analytic up-to-horizon~\cite{BeigChAnalyticity} near the horizon, but this is irrelevant for the problem here.) For this, we recall some standard facts about isometries. Let us denote by
\newcommand{\bifurc}{{\mycal Z}}%
\bel{3IV17.1}
 \bifurc:=\{0\}\times  \ourN
\ee
the codimension-two submanifold of $\Mriem$  which is the zero-set of the Killing vector  $X$ generating rotations of $\R^2$. Then $\bifurc$ is a totally-geodesic submanifold of $(\Mriem,\Re \griem)$. In coordinates $(x,y)$ normal for the metric $\Re \griem$, on each of the planes $\Re\griem$--orthogonal to $\bifurc$ the  vector field $X$ takes the standard Euclidean form
$$
 X = x\partial_y - y \partial_x
 \,.
$$
This shows that in these coordinates a rotation $R_\pi$ by an angle $\pi$, which is the map $(x,y)\mapsto (-x,-y)$, is an isometry of $\Re \griem$ which leaves invariant $\Im \griem$.
 Let us choose local coordinates $(x^a)$ on $\bifurc$, and extend them to be constant along $\Re\griem$--geodesics $\Re\griem$--normal to $\bifurc$. We will denote by $(x^A)$ the coordinates $(x,y)$. One obtains
$$
 \Re\griem_{ab}(x,y,x^c) = \Re\griem_{ab}(-x,-y,x^c)
 \,,
 \quad
 \Re\griem_{AB}(x,y,x^c) = \Re\griem_{AB}(-x,-y,x^c)
 \,,
$$
$$
 \Re\griem_{aA}(x,y,x^c) = -\Re\griem_{aA}(-x,-y,x^c)
 \,.
$$
In particular all odd-order derivatives of the metric functions
$ \Re\griem_{ab}$ and $\Re\griem_{AB} $  vanish on $\bifurc$.

An analogous argument applies to $\theta$ using $\mcL_X \theta =0$.

Let us assume for definiteness that $a\in\R$, thus $\theta$ is purely real. It is then standard to derive the following form of the metric in coordinates $(\varphi,\rho,x^a)$, where $(x,y)=(\rho\cos\varphi, \rho\sin\varphi)$ (compare~\cite[Section~3]{ChAPP} for detailed calculations in a closely related setting):
\beal{3IV17.2}
 &
\Re\griem = u^{2}  d\varphi  ^{2} + h_{jk}dx^j dx^k\,,
 \quad
 u = \rho(1+\rho^2 \psi)
 \,,
 &
\\
 &
 \theta  = \frac {\alpha \rho }{ (1+\rho^2\psi)^2}d \rho  +
 \gamma_a dx^a\,,
  &
\\
 &    \quad
 %\phantom{x}
h_{jk}dx^j dx^k=
(1+\rho^2\beta)  d \rho^2 +
b_{ab}dx^a dx^b + 2 \rho \lambda_a dx^a d \rho- u^2 \theta_i \theta_j
dx^i dx^j \,,
 &
\eeal{3IV17.3}
and where all the non-explicit functions are smooth functions of $(\rho^2,x^a)$.

Passing to the Lorentzian regime, and replacing $\varphi$ by a coordinate
\bel{4IV17.1}
 \tau = \varphi + \log \rho
 \,,
\ee
one checks that the Lorentzian metric $\glorentz$ smoothly extends to a Killing horizon at $\rho=0$ after a final change of coordinates $\rho\to z=\frac 12 \rho^2$. Indeed, the Lorentzian metric $\glorentz$ is then given by
\bel{5IV17.1}
 \begin{split}
\glorentz=&-u^{2}  (d\tau^2-\frac{2}{\rho} d\tau d\rho)+\rho^2 (\beta - 2\psi-\rho^2\psi^2)d\rho^2 \\
&+ b_{ab}dx^a dx^b + 2\rho\lambda_a dx^a d\rho -u^2 \theta_i\theta_j dx^i dx^j
\\ =&-2(1+\rho^2 \psi)^2  (   z\,d\tau^2-  d\tau dz)+  (\beta - 2\psi-\rho^2\psi^2)dz^2 \\
&+ b_{ab}dx^a dx^b + 2\lambda_a dx^a dz -2z(1+\rho^2 \psi)^2  \theta_i\theta_j dx^i dx^j\,,
 \end{split}
 \ee
after substituting $d{\Phi}\mapsto i d{\Phi}$ in \eqref{3IV17.2} and applying the coordinate transformation \eqref{4IV17.1}.

Indeed, \eq{5IV17.1} shows that that the Killing vector $\partial_\tau$ is null on the hypersurface $\{z=0\}$, and that this hypersurface is null, hence a Killing horizon. This is a \emph{non-rotating} horizon, in the sense that the Killing vector which is timelike at infinity is also tangent to the Killing horizon. (This explains why our solutions, which can have no further symmetries than stationarity, are compatible with the Hollands-Ishibashi-Wald~\cite{HIW} rigidity theorem, which provides at least one more symmetry for rotating horizons.) We also see from \eq{5IV17.1} that the surface gravity of the Killing horizon $\{z=0\}$, calculated for the vector field $\partial_\tau$, equals one. Rescaling $\tau$ to the scale of the original nearby seed Birmingham solution, the surface gravity of our solutions will coincide with that of the seed metric in those cases with matter sources where the gravitational free data at infinity have been chosen to coincide with the original ones; otherwise a nearby surface gravity will result when the asymptotic behaviour of the metric imposes a natural rescaling of the horizon Killing vector field.

\bigskip

\noindent{\sc Acknowledgements} The research of PTC was supported in
part by the Austrian Science Fund (FWF), Projects  P23719-N16 and P29517-N27, and by the Polish National Center of Science (NCN) under grant 2016/21/B/ST1/00940. PK was supported by a uni:docs grant of the University of Vienna. We are grateful to the Erwin Schr\"odinger Institute for hospitality and support during part of work on this paper.

\bibliographystyle{amsplain}

\bibliography{EMDBlackHoles-minimal}

\def\polhk#1{\setbox0=\hbox{#1}{\ooalign{\hidewidth
  \lower1.5ex\hbox{`}\hidewidth\crcr\unhbox0}}} \def\cprime{$'$}
  \def\cprime{$'$}
\providecommand{\bysame}{\leavevmode\hbox to3em{\hrulefill}\thinspace}
\providecommand{\MR}{\relax\ifhmode\unskip\space\fi MR }
% \MRhref is called by the amsart/book/proc definition of \MR.
\providecommand{\MRhref}[2]{%
  \href{http://www.ams.org/mathscinet-getitem?mr=#1}{#2}
}
\providecommand{\href}[2]{#2}
\begin{thebibliography}{10}

\bibitem{mand2}
M.T. Anderson, \emph{{Einstein} metrics with prescribed conformal infinity on
  $4$-manifolds}, Geom. Funct. Anal. \textbf{18} (2001), 305--366,
  arXiv:math.DG/0105243. \MR{2421542}

\bibitem{mand1}
\bysame, \emph{Boundary regularity, uniqueness and non-uniqueness for {AH
  Einstein} metrics on $4$-manifolds}, Adv.\ in Math. \textbf{179} (2003),
  205--249, arXiv:math.DG/0104171. \MR{2010802}

\bibitem{ACD}
M.T. Anderson, P.T. Chru\'{s}ciel, and E.~Delay, \emph{Non-trivial, static,
  geodesically complete vacuum space-times with a negative cosmological
  constant}, Jour.\ High Energy Phys. \textbf{10} (2002), 063, 22 pp.,
  arXiv:gr-qc/0211006. \MR{1951922}

\bibitem{ACD2}
\bysame, \emph{Non-trivial, static, geodesically complete space-times with a
  negative cosmological constant. {II}. {$n\geq5$}}, AdS/CFT correspondence:
  Einstein metrics and their conformal boundaries, IRMA Lect. Math. Theor.
  Phys., vol.~8, Eur. Math. Soc., Z\"urich, 2005, arXiv:gr-qc/0401081,
  pp.~165--204. \MR{MR2160871}

\bibitem{AstefaneseiRadu}
D.~Astefanesei and E.~Radu, \emph{Boson stars with negative cosmological
  constant}, Nucl.\ Phys.\ B \textbf{665} (2003), 594--622,
  arXiv:gr-qc/0309131. \MR{2000918}

\bibitem{BeigChAnalyticity}
R.~Beig and P.T. Chru\'{s}ciel, \emph{{{On analyticity of stationary vacuum
  metrics at non-degenerate Killing horizons}}}, in preparation.

\bibitem{BiquardContinuation}
O.~Biquard, \emph{{M\'etriques d'Einstein asymptotiquement sym\'etriques
  (Asymptotically symmetric Einstein metrics)}}, {Ast\'erisque 265, Paris:
  Soci\'et\'e Math\'ematique de France, 109 pp.\ }, 2000.

\bibitem{Birmingham}
D.~Birmingham, \emph{Topological black holes in anti-de {Sitter} space},
  Class.\ Quantum Grav. \textbf{16} (1999), 1197--1205, arXiv:hep-th/9808032.
  \MR{MR1696149 (2000c:83062)}

\bibitem{BjorakerHosotani}
J.~Bjoraker and Y.~Hosotani, \emph{{Monopoles, dyons and black holes in the
  four-dimensional Einstein-Yang-Mills theory}}, Phys.\ Rev. \textbf{D62}
  (2000), 043513, arXiv:hep-th/0002098.

\bibitem{BBSLMR}
J.L. Bl\'azquez-Salcedo, J.~Kunz, F.~Navarro-L\'erida, and E.~Radu,
  \emph{{Static Einstein-Maxwell Magnetic Solitons and Black Holes in an Odd
  Dimensional AdS Spacetime}}, Entropy \textbf{18} (2016), 438,
  arXiv:1612.03747 [gr-qc].

\bibitem{ChernSimons}
S.S. Chern and J.~Simons, \emph{Characteristic forms and geometric invariants},
  Ann. of Math. (2) \textbf{99} (1974), 48--69. \MR{0353327}

\bibitem{ChAPP}
P.T. Chru\'{s}ciel, \emph{On analyticity of static vacuum metrics at
  non-degenerate horizons}, Acta Phys.\ Pol. \textbf{B36} (2005), 17--26,
  arXiv:gr-qc/0402087.

\bibitem{ChDelayStationary}
P.T. Chru\'{s}ciel and E.~Delay, \emph{Non-singular, vacuum, stationary
  space-times with a negative cosmological constant}, Ann. Henri Poincar\'e
  \textbf{8} (2007), 219--239. \MR{MR2314449}

\bibitem{ChDelayEM}
P.T. Chru{\'s}ciel and E.~Delay, \emph{{Non-singular spacetimes with a negative
  cosmological constant: II. Static solutions of the Einstein-Maxwell
  equations}},  (2016), arXiv:1612.00281 [math.DG].

\bibitem{ChDelayKlinger}
P.T. Chru{\'s}ciel, E.~Delay, and P.~Klinger, \emph{{Non-singular spacetimes
  with a negative cosmological constant: III. Stationary solutions with matter
  fields}},  (2017).

\bibitem{ChDelayKlingerNonDegenerate}
\bysame, \emph{{On non-degeneracy of Riemannian Schwarzschild-anti de Sitter
  metrics}},  (2017), arXiv:1710.07597 [gr-qc].

\bibitem{ChDelayKlingerBoson}
P.T. Chru{\'s}ciel, E.~Delay, P.~Klinger, A.~Kriegl, P.W. Michor, and
  A.~Rainer, \emph{{Non-singular spacetimes with a negative cosmological
  constant: V. Boson stars}},  (2017).

\bibitem{DeFelice2010}
Antonio De~Felice and Shinji Tsujikawa, \emph{f(r) theories}, Living Reviews in
  Relativity \textbf{13} (2010), no.~1, 3.

\bibitem{DiasHorowitzSantos}
O.J.C. Dias, G.T. Horowitz, and J.E. Santos, \emph{{Black holes with only one
  Killing field}}, Jour. High Energy Phys. (2011), 115, 43, arXiv:1105.4167
  [hep-th]. \MR{2875937}

\bibitem{GL}
C.R. Graham and J.M. Lee, \emph{{Einstein} metrics with prescribed conformal
  infinity on the ball}, Adv.\ Math. \textbf{87} (1991), 186--225.

\bibitem{Guillarmou}
C.~Guillarmou, \emph{Meromorphic properties of the resolvent on asymptotically
  hyperbolic manifolds}, Duke Math.\ Jour. \textbf{129} (2005), 1--37.
  \MR{2153454}

\bibitem{HT}
M.~Henneaux and C.~Teitelboim, \emph{Asymptotically anti--de {S}itter spaces},
  Commun.\ Math.\ Phys. \textbf{98} (1985), 391--424. \MR{86f:83030}

\bibitem{HIW}
S.~Hollands, A.~Ishibashi, and R.M. Wald, \emph{A higher dimensional stationary
  rotating black hole must be axisymmetric}, Commun.\ Math. Phys. \textbf{271}
  (2007), 699--722, arXiv:gr-qc/0605106.

\bibitem{lee:spectrum}
J.M. Lee, \emph{The spectrum of an asymptotically hyperbolic {E}instein
  manifold}, Comm.\ Anal.\ Geom. \textbf{3} (1995), 253--271.

\bibitem{Lee:fredholm}
\bysame, \emph{Fredholm operators and {E}instein metrics on conformally compact
  manifolds}, Mem. Amer. Math. Soc. \textbf{183} (2006), vi+83,
  arXiv:math.DG/0105046. \MR{MR2252687}

\bibitem{MazzeoMelrose}
R.R. Mazzeo and R.B. Melrose, \emph{Meromorphic extension of the resolvent on
  complete spaces with asymptotically constant negative curvature}, Jour.\
  Funct.\ Anal. \textbf{75} (1987), 260--310. \MR{916753}

\bibitem{OkuyamaMaeda}
N.~Okuyama and K.~Maeda, \emph{{Five-dimensional black hole and particle
  solution with nonAbelian gauge field}}, Phys. Rev. \textbf{D67} (2003),
  104012, arXiv:gr-qc/0212022 [gr-qc].

\bibitem{RaduTchrakian}
Eugen Radu and D.~H. Tchrakian, \emph{{Gravitating Yang-Mills fields in all
  dimensions}}, {418th WE-Heraeus-Seminar: Models of Gravity in Higher
  Dimensions: From theory to Experimental search Bremen, Germany, August 25-29,
  2008}, 2009, arXiv:0907.1452 [gr-qc].

\bibitem{WinstanleyEYM}
E.~Winstanley, \emph{{Existence of stable hairy black holes in SU(2) Einstein
  Yang-Mills theory with a negative cosmological constant}}, Class.Quant.Grav.
  \textbf{16} (1999), 1963--1978.

\end{thebibliography}
\end{document}